# Transmission of Successful Route Error Message(RERR) in Routing Aware Multiple Description Video Coding over Mobile Ad-Hoc Network

Kinjal Shah[1],Gagan Dua[1], Dharmendar Sharma[1], Priyanka Mishra[1], Nitin Rakesh[1]

[1]Department of Computer Science &Engineering, Jaypee University, Waknaghat, Dist.Solan (H.P)

`kinjal.93@gmail.com, d.sharma000@gmail.com, dua.gagan@live.in, mishrapriyanka6@gmail.com,nitin.rakesh@gmail.com`

## ABSTRACT

*Video transmission over mobile ad-hoc networks is becoming important as these networks become more widely used in the wireless networks. We propose a routing-aware multiple description video coding approach to support video transmission over mobile ad-hoc networks with single and multiple path transport. We build a model to estimate the packet loss probability of each packet transmitted over the network based on the standard ad-hoc routing messages and network parameters without losing the RERR message. We then calculate the frame loss probability in order to eliminate error without any loss of data.*

## 1. INTRODUCTION

There has been a growing interest in video communications over mobile ad-hoc networks because of the ease of communication between the two terminal nodes due to this ad-hoc networks is used in various applications like military, homeland defense, and disaster recovery. However, mobile ad-hoc networks face various challenges to video transmissions because mobility of nodes the lack of infrastructure in the network can lead to the link failures and route changes.

In previous case, multiple description video coding (MDC), when any intermediate node is failed in transmitting the packet due to link failure then it will send a message in the form of RERR to the source node in ad hoc networks. When failure is noticed by any of intermediate nodes then the immediate intermediate node will send the information of failure to its previous node and this node will send a message to previous node in the form of RERR message, and will be transmitted to the source, but if the RERR message is being transferred from one node to another node then it may possible that link between intermediate node & transmitter may get failed and this causes the failure of RERR message. Here we are proposing the solution transmission of data without losing the RERR message.

We will propose this model for both unicast and multipath transport. For this each intermediate node should have one routing table with key values of source address and destination address in pair. With the help of address information each node will have knowledge about from where the packets are coming and where it needs to be sent the packet. In case of unicast network we have single link for sending & receiving data from one node to another node. Each and every node including transmitter & receiver will have certain time stamps so they will transmit the video packets if any failure occurs. And we will propose the same solution for the multicast network.

We present how to use routing messages to estimate the packet losses in the network. Based on the routing mechanisms, a route error (RERR) message is initiated .This RERR indicates that a link becomes unreliable and the packets transmitted through this link suffer a high packet loss rate. Due to the random delay between link failure and RERR reception at the source, we use two states to represent these two cases are: GOOD means the packet is sent before the link failure, FAIL means the packet fails to transmit and triggers RERR .With the help of this analysis we will calculate the packet loss probability and also the probability of frame corruptionfinally, we will make use of algorithm, which simplifies the reference frame selection.

## 2. RELATED WORK

Uptill now in this area of research one area goes towards how effective MDC methods are with path diversity [4-6].An adaptive mode approach focuses to adapt for network condition as well as video characteristics. This approach selects MD mode by calculating end to end distortion based on Gilbert packet loss model[1].In [5] Mao has proposed MDC schemes with multipart transport & noted that MDC is preferable even when a feedback channel can't be set up.

Another area goes towards path selection & rate allocation for MDC given a particular MDC scheme [10]-[13].Begen et.al proposed a multipath selection method which chooses a set of paths maximizing the overall quality of client based on the network parameters [10].In [12] the author formulated the video distortion as a function of network layer behaviour & proposed a branch & bound framework to generate optimal solutions.In [1] author considered a multipath routing in a more practical network & utilizes the route messages to select the proper reference frame.Our work is inspired by the ACK/NACK for every video packet [14].

In our approach we used source & destination address being maintained at each intermediate node & if receiver got the packet successfully & getting duplicate packets then according to our approach receiver sends the ACK,NACK packets to its previous nodes respectively, such an approach can lead to extra overhead & cost specifically in a large network having many much nodes in which doesn't require any additional control packet or an extra channel in the network.

## 3. SYSTEM ARCHITECTURE

In previous case of multiple description video coding for ad hoc network when any intermediate node gets failure in transmitting the packet due to link failure then it will try to inform the source node as early as possible time. Now when failure is noticed by any of intermediate nodes then the current intermediate node will inform its previous node and from its previous node the RERR message will get transmitted to the source, but if the RERR message is being transferred from one node to another node then it may possible that link between intermediate node & transmitter may get failed at this time the RERR message may get lost, So here we are losing the RERR message. Here we are proposing the solution for not losing the RERR message.

## 3.1. Proposed Model

Our proposed model for sending successful RERR message from intermediate node to transmitter is that each intermediate node should have one routing table having key values of source address & destination address in pair. (Source address, Destination address). Using the address information every node is having knowledge about from where this packet had come & where it need to send the packet further in case of unicast network scenario where there is only one link for sending & receiving data from one node to another node till the packet of video reach to the receiver finally. Each & every node including transmitter & receiver will have certain time stamps so they will transmit the video packets if any failure occurs. Based on time stamp they will retransmit the packet if no acknowledgement comes.

We have designed our model in such a way when one node will send the video packet to the next its neighbor intermediate node & waits till itsacknowledgement (ACK) come in specific time stamp so it can understand that there is no link failure as the packet has come in time so it can inform to transmitter with in specific time stamp so it can send the further packet s to the first intermediate node. If receiver gets the video packet the from intermediate nodes then it will send the acknowledgement to previous node but if acknowledgement fails due to noise then second time it will send NACK signal which means Negative Acknowledgement for preventing the nodes to transmit the duplicate packets again & again.

We have explained the scenario in the diagram as shown below.

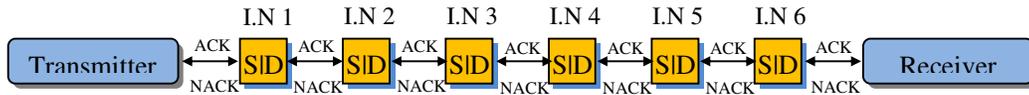

Figure 1.Unicast Network having one transmitter & receiver

In above diagram N1, N2,Nk,Nk+1 indicates the intermediate nodes between the source & receiver & SID indicates that each intermediate node is having routing table having Source Address & Destination Address with specific time stamp

## 3.2. Packet Loss Estimation for Proposed Model

In our proposed model the video packet would be in two states Good & Failure. We are denoting probability of good packets as$\lambda(g)$ & $\lambda(f)$ [1]. $Pg(n)$ & $Pf(n)$ are probability of the $n$th preceding packets in these two states [1]. The Equation is as follows. $Pr(n)$is the probability for packet Loss estimation [1].

$$Pr(n) = \lambda g \times Pg(n) + \lambda f \times Pf(n). \qquad [1]$$

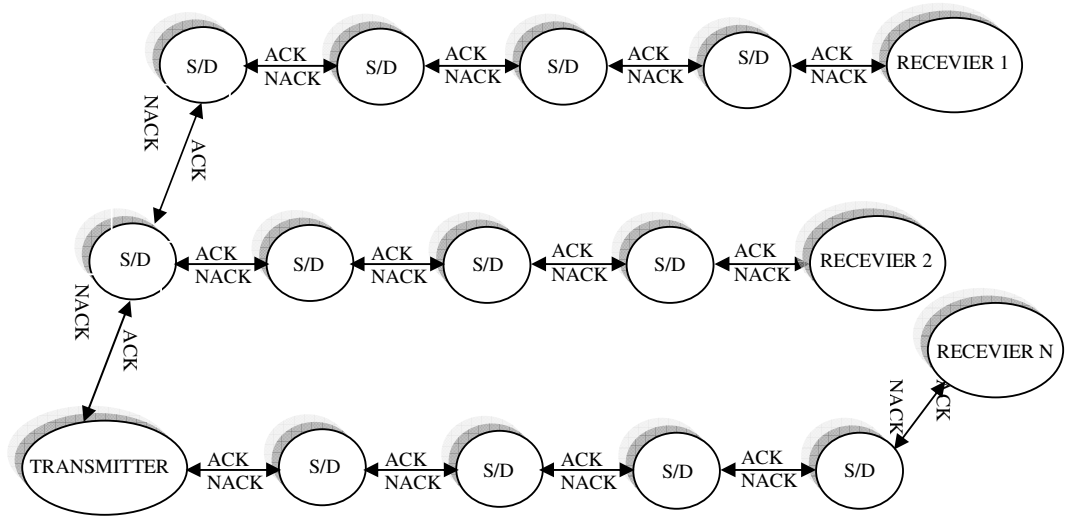

Figure 2.Multicast Network having one Transmitter & Multiple Receivers

### 3.3. Estimation of State Probability Distribution

In our proposed model as our video packets are having just two states Good & Failure so finally we can say indisputably the packets of states. We need to calculate the probability for the $Pg(n)$ as well as $Pf(n)$.

$$Pg(n) = p(T_{delay} <= (n-1) \times T_{data})$$

$$Pf(n) = p(n \times T_{data} >= T_{delay} > (n-1) \times T_{data})$$

We can calculate the video packet interval data rate by $T_{data} = \frac{L}{R_t}$.In our proposed model the video packet would be in two states Good & Failure. We are denoting probability of good packets as $\lambda(g)$ & $\lambda(f)$ [1]. $P_g(n)$ & $P_f(n)$are probability of the nth preceding packets in these where L is payload size & $R_t$ is the transmission rate [1].So our Delay would be equated as follows $T_{delay} = 2 \times T_{retrans} + T_{RERR}$ where $T_{retrans}$ .We can calculate the video packet interval data rate by $T_{data} = \frac{L}{R_t}$ is the time for retransmission & $T_{RERR}$ is route error message one $T_{retrans}$ for sending the packet & another $T_{retrans}$ for receiving acknowledgement.

### 3.4. Probability of Corrupted Frame

$p(f_k) = 1 - \pi v_i \in f_k(1 - p(v_i))$ where $f_k$ is the probability of frame corruption, $p(v_i)$ is the packet loss probability for packet $v_i$ in frame $f_k$.

Table 1.Proposed Algorithm.

| Step 1 | While have video context to send do |
|---|---|
| Step 2 | While have video context to send do |
| Step 3 | For all reference frames in current description do |
| Step 4 | If the frame corruption probability of the reference is larger than a threshold the |
| Step 5 | Remove the frame from the reference frame list |
| Step 6 | End if |
| Step 7 | End for |
| Step 8 | If no frames are available in the reference list then |
| Step 9 | Add available frames in the other description to the reference list |
| Step 10 | End if |
| Step 11 | Encode a packet of video using selected references and transmit it through one of the two path |
| Step 12 | If Intermediate node receives an RERR message that implies a link failure then |
| Step 13 | Send the RERR message to the previous neighbour node based on source address stored in the routing table |
| Step 14 | If  RERR message is lost before reaching to the transmitter then |
| Step 15 | Forward RERR message to the transmitter based on destination address of previous node |
| Step 16 | Else continue forwarding RERR signal till it reaches to the transmitter |
| Step 17 | End if |
| Step 18 | Else |
| Step 19 | If intermediate node gets the packet successfully then |
| Step 20 | Send ACK signal to previous intermediate node |
| Step 21 | Else Retransmit the packet till the ACK comes within specific time stamp. |
| Step 22 | End if |
| Step 23 | If receiver is getting duplicate packets then send NACK signal to the sender |
| Step 24 | Else send ACK signal to the sender. |
| Step 25 | End if |
| Step 26 | Estimate the RERR delay and determine the packet loss probability for affected packets. |
| Step 27 | If route is available in the route cache then |
| Step 28 | Reconstruct a new route from the route cache |
| Step 29 | Else |
| Step 30 | Initiate route recovery process |
| Step 31 | Repeat |
| Step 32 | Mark the packets scheduled to be sent through the broken route as lost |
| Step 33 | Until receive an RREP to build a new route |
| Step 34 | End if |
| Step 35 | Estimate the frame corruption probability based on the estimated packet loss probability |
| Step 36 | End if |
| Step 37 | End if |
| Step 38 | End while |

In the above algorithm we have modified steps from 12 to 25 for better transmission of RERR message to the transmitter & thus we can improve the efficiency of the transmitter by removing the redundancy for resending the video packets.

## 3.5. Advantage of proposed Model

In previous model when intermediate node finds the failure & wants to send the RERR message to the previous node & from previous node to transmitter but if before reaching RERR message to the transmitter if the link fails then there was no way for transmitter for getting RERR message. In our model as each & every intermediate node is having source address & destination address so if any link fails during the transmission of RERR message towards transmitter from the destination address of previous node, it can find out the link failure in the network so it can inform the transmitter. In our case in any circumstances the RERR signal will never get lost.

With the help of negative acknowledgement signal NACK any node can get notification regarding the packet has reached successfully towards the receiver so unnecessary retransmission can be prevented.In Unicast network as well as Multicast Network this proposed model will work in very efficient manner because network will not get overloaded so load balancing problem will be solved automatically.

Using this approach we are getting confirm delivery of video packets with constant time delay of $T_d$. Even if receiver receives the packets after delay $T_d$ but no packet would get lost.

## 3.5. Limitation

This approach is having multiple intermediate nodes between transmitter & receiver so every node between sender & receiver must be up during the time of transmission.for a multicast network there are multiple receivers then sometimes there can be more delay in transmission of video packets due to there is one sender which has to put the video packets on the network & that packet comes via passing intermediate nodes, sometimes sender takes much time on putting video packets on the link due to heavy demand from the receiver side & sender gets bottlenecked.Delay can be much more than expected when there are large no. of video packets to be transmitted for multiple users then getting ACK signal at each intermediate node for transmission of next video packets can also take plenty of time so system can be slower in this case. In this case Multipath routing is more suitable instead of unicast or multicast network because they provide just one link for sending packets as well as receiving ACK signals.

## 4. Conclusion

we proposed the transmission model for successful transmission of RERR message In Routing Aware MDC for video over mobile ad-hoc network from intermediate nodes to their previous neighbor source node & thus via source node finally RERR message successfully reaches towards its main sender. Using this approach we can say that we are getting delayed but guaranteed video packets at the receiver side. Video quality will be guaranteed as no packet can get lost from the network & bandwidth of the network can also be saved by avoiding retransmission of packets without knowing whether further link is good for transmission or damaged because in our proposed model every intermediate nodes as well as transmitter will send the further packets after getting ACK signal for the previous sent packets.

## Authors

**Kinjal Shah** received his B.E degree in Computer Engineering from Arvindbhai.D.Patel Institute of Technology,NewVallabhVidyanagar, Distt.Anand in 2009. He is currently pursuing MTECH degree in Computer Science & Engineering at Jaypee University of Information Technology,Waknaghat,Solan-173215. His research interest includes Cloud Computing, Cryptography & Network Security, Multimedia Data Transmission in ad-hoc networks.

**GaganDua** received his B.E. degree in Computer Engineering from ShriKrishan Institute of Engineering & Technology, Kurukshetra in 2009. He completed his three years diploma in Computer Engineerign from Seth Jai Parkash Polytechnic, Damla (Yamunanagar, Haryana).Currently he is pursuing MTECH degree in Computer Engineering at Jaypee University of Information Technology, waknaghat, Solan-173215. His areas of interest are multimedia data communication & microprocessor.

**Dharmendar Sharma** received his B.E degree in Information Technology from Ajay Kumar Garg Engineering College, Ghaziabad in 2009. Currently he is pursuing MTECH degree in Computer Science & Engineering at JaypeeUniversity of Information Technology, waknaghat, Solan-173215. His area of interests are data warehousing & data Mining, data communication & networking.

**Priyanka Mishra** received her B.E degree in Electronics & Communication Engineering from BBS college of Engineering & Technology, Allahabad in 2009. Currently she is pursuing MTECH degree in Electronics & Communication Engineering at Jaypee University of Information Technology, waknaghat, Solan-173215. Her area of interests are data communication & networking.

**NitinRakesh** is Sr. Lecturer in the Department of Computer Science and Engineering & Information Technology, Jaypee University of Information Technology (JUIT), Waknaghat, Solan–173215, Himachal Pradesh, India. In 2004, he received the Bachelor's Degree in Information Technology and Master's Degree in Computer Science and Engineering in year 2007. Currently he is pursuing his doctorate in Computer Science and Engineering and his topic of research is parallel and distributed systems. He is a member of IEEE, IAENG and is actively involved in research publication. His research interest includes Interconnection Networks & Architecture, Fault–tolerance & Reliability, Networks–on–Chip, Systems–on–Chip, and Networks–in–Packages, Network Algorithmics, Parallel Algorithms, Fraud Detection. Currently he is working on Efficient Parallel Algorithms for advanced parallel architectures.